**Magnetic inhibition of convection in O star envelopes**

James MacDonald & Véronique Petit
*Department of Physics and Astronomy, University of Delaware, Newark, DE 19716, USA*

**ABSTRACT**

It has been suggested that the absence of macroturbulence in the atmosphere of NGC 1624 – 2 is due its strong magnetic field (the strongest known for a massive O star) suppressing convection in its outer layers, removing the mechanism thought responsible for the observed macroturbulence in stars with lower field strengths. Here, we develop and apply a criterion for a uniform magnetic field to suppress convection in stellar envelopes in which radiation pressure is a significant contributor to hydrostatic balance. We find upper mass limits of ~ 55 $M_\odot$ and ~ 30 $M_\odot$ for magnetic suppression to be possible in Zero Age Main Sequence and Terminal Age Main Sequence stars, respectively. For evolved stars, magnetic suppression of convection can significantly alter the stars' evolution. For NGC 1624 – 2, we find a polar dipole strength of 16.5 ± 5.9 kG is required to suppress convection, in good agreement with the value ~20 kG measured by spectropolarimetry.

## 1 INTRODUCTION

Doppler broadening of spectral lines from the photospheres of OB stars is attributed to a combination of rotation, outflows and macroturbulence. The origin of macroturbulence is poorly understood. Cantiello et al. (2009) investigated the observational consequences of the thin near-surface convection zone that results from iron-bump opacity in the outer envelope of a hot star. Based on stellar structure models to determine the convection zone properties, they argue for a physical connection between the sub-photospheric convective motions and small scale stochastic velocities in the photosphere of O- and B-type stars, often interpreted as microturbulence and/or macroturbulence. Saio et al. (2006) showed that excitation of g-mode pulsations by iron-bump opacity could occur in post main sequence slowly pulsating B supergiants. Aerts et al. (2009) proposed that macroturbulence in such stars results from the collective effect of a large number of low-amplitude non-radial gravity-mode pulsations. Aerts & Rogers (2015) have suggested that stochastically excited g-modes are responsible for O-star macroturbulence. Sundqvist et al. (2013, hereafter S13) tested the hypothesis that g – mode oscillations, excited by the iron-bump opacity convection zone in the outer envelope of a hot star, is the origin of the observed turbulent broadening by examining macroturbulence in eight slowly rotating magnetic O stars.

Magnetic measurements offer an unambiguous way of measuring the rotation period of magnetic OB stars (Stibbs 1950; Monaghan 1973; Shultz & Wade 2017). With rotational periods ranging from 7 days to ~50 years, the line broadening in these 8 magnetic O-type stars is expected to be dominated by macroturbulence. S13 find that this is the case except for NGC 1624-2, which is the star in the sample with by far the strongest magnetic field. Instead, NGC 1624 - 2 has very sharp CIV lines dominated by magnetic Zeeman broadening (~20 km s$^{-1}$), which is negligible in the other stars in the sample. S13 hypothesize that the generation of macroturbulence is suppressed in this star because the strong magnetic field stabilizes the atmosphere. To determine whether a sufficiently strong magnetic field can suppress convection associated with the iron bump, they use a simple model based on the idea that convection is suppressed when the magnetic pressure is comparable to the thermal pressure in the convection zone.

In section 2, we point out a number of problems with the approach of S13 and provide an alternative criterion to determine the critical field strength needed to suppress convection, based on a rigorous analysis of convective instability for ideal MHD conditions that allows for deviations from an ideal gas equation of state. In section 3, we first use our non-magnetic stellar models of O stars to determine the critical field strengths to suppress convection in main sequence stars. In addition to convection driven by Fe - bump opacity, we also consider convection due to He+ ionization that can occur in the envelopes of the cooler O stars. We then include in our stellar evolution models a model for magnetoconvection that is based on our instability criterion, with the goal of determining the critical field needed to completely suppress convection in a stellar model at a particular point of its main sequence evolution. Our conclusions and discussion are given in section 4.

## 2 ESTIMATES OF CRITICAL FIELD STRENGTHS

### 2.1 Model of Sundqvist et al. (2013)

In developing their simple model for suppression of convection by a magnetic field, S13 make a number of assumptions. They assume that the Rosseland mean opacity, $\kappa$, is uniform, which allows the thermal pressure, $p$, to be related to optical depth, $\tau$, by

$$p\kappa = g\tau. \tag{1}$$

S13 refer to the pressure as the gas pressure but in this equation, which is derived using the equation of hydrostatic equilibrium, $p$ is actually the total pressure, including radiation pressure. They adopt an Eddington-approximation temperature structure for a radiative grey atmosphere

$$T^4 = T_{eff}^{\;4}\left(\frac{3}{4}\tau + \frac{1}{2}\right). \tag{2}$$

They assume that convection is suppressed when the magnetic pressure equals the thermal pressure, independent of the direction of the magnetic field. These assumptions lead to their critical field estimate

$$B_S = \sqrt{\frac{32\pi}{3}\frac{g}{\kappa}}\left(\frac{T}{T_{eff}}\right)^2. \tag{3}$$

There are a number of difficulties with this approach. The major one being that is independent of any aspects of convective instability in the presence of magnetic field, a subject that has a long history of study. Another difficulty is that it does not make clear how the magnetic field interacts with the contribution of radiation to the thermal pressure. Also, the Rosseland mean opacity is not uniform (a fully ionized, uniform opacity envelope would not be convective even in the absence of a magnetic field) but, e.g. for a 30 $M_\odot$ ZAMS model, varies from order unity at low optical depths to ~ 2 – 3 cm$^2$/g at the location of the Fe opacity bump, which is the main driver for convective instability.

## 2.2 Estimate of critical field strength from mixing length theory for convection

Here we estimate the magnetic field strength needed to suppress convection by comparing the kinetic energy density of convective motions to the magnetic energy density. In the mixing length theory of convection (in the absence of magnetic field), the convective velocity, $v$, is given by

$$v^2 = f_1 Q \frac{\rho}{p} g^2 l^2 (\nabla - \nabla_{ad}), \qquad (4)$$

where $\rho$ is the density, $p$ is pressure, $g$ is gravitational acceleration, $l$ is the mixing length, $\nabla$ is the structural gradient, $\nabla_{ad}$ is the adiabatic gradient, and $f_1$ is a constant of order unity. The remaining quantity, $Q$, is a thermal expansion coefficient given by

$$Q = -\left(\frac{\partial \ln \rho}{\partial \ln T}\right)_p, \qquad (5)$$

where $T$ is temperature. For ideal gas plus radiation, the exact expression is

$$Q = 1 + 4 \frac{p_{rad}}{p_{gas}}, \qquad (6)$$

where $p_{gas}$ and $p_{rad}$ are the gas and radiation pressures, respectively. In the envelopes of massive O stars, radiation pressure is at least as important as gas pressure and $Q$ can be significantly greater than unity.

By making the usual assumption that the mixing length is of the same order as the pressure scale height, the ratio of magnetic energy density to kinetic energy density of convective motions is

$$\frac{B^2}{4\pi\rho v^2} = \frac{B^2}{4\pi\rho} \frac{\rho}{p} \frac{1}{f_2 Q (\nabla - \nabla_{ad})}, \qquad (7)$$

where $f_2$ is a second constant of order unity. This ratio is order unity when

$$\frac{B^2}{4\pi\rho} \sim \frac{p}{\rho} Q (\nabla - \nabla_{ad}). \qquad (8)$$

This relation can also be interpreted in terms of Alfven velocity, $v_A$, and sound speed, $c_s$. It indicates that magnetic suppression of convection occurs when

$$\frac{v_A^2}{c_s^2} \sim Q (\nabla - \nabla_{ad}). \qquad (9)$$

## 2.3 Estimate of critical field strength from magneto-convective stability criteria

A rigorous derivation of a relation similar to that in equation (9) has been given by Gough & Tayler (1966) for the case of an ideal gas, for which $Q = 1$. Gough & Tayler used an energy principle (Bernstein et al. 1958) to show that a sufficient condition for stability of a perfectly conducting ideal gas against convection in a uniform *vertical* field is

$$\nabla - \nabla_{ad} < \frac{B^2}{B^2 + 4\pi\gamma p} = \frac{v_A^2}{v_A^2 + c_s^2}, \tag{10}$$

where $\gamma$ is the ratio of specific heats. Earlier work (Newcomb 1961; Tayler 1961) had already shown that a uniform horizontal field does not change the Schwarzschild criterion for convection.

Moreno – Insertis & Spruit (1989; hereafter MIS89) used a linear stability analysis of the ideal MHD equations to generalize the Gough – Tayler criterion for stability to allow for variations of the molecular weight, $\mu$.

MacDonald & Mullan (2009) modified the Gough – Tayler criterion to allow for non-ideal gas behavior by multiplying the superadiabaticity by the thermal expansion coefficient, $Q$. The specific form of their criterion for stability is

$$Q(\nabla - \nabla_{ad}) < \frac{v_A^2}{v_A^2 + c_s^2}. \tag{11}$$

For weak enough fields, this criterion is equivalent to that in equation (9), provided that only the vertical component of the field is considered to be responsible for magnetic suppression of convection.

An important property of equation (11) is that the right hand side is bounded above by unity. Since $Q$ can become quite large in massive stars, the possibility arises that a magnetic field is unable to suppress convection, no matter its strength.

A more detailed analysis given in the Appendix shows that for a general equation of state the criterion for stability is

$$Q(\nabla - \nabla_{ad}) - \frac{v_A^2}{v_A^2 + c_s^2}\left(1 + \frac{d\ln\Gamma_1}{d\ln p}\right) < 0, \tag{12}$$

where $\Gamma_1$ is the first adiabatic exponent (the partial derivative of $\log p$ with respect to $\log \rho$ at constant entropy). For a mixture of ideal gas and radiation

$$Q = \frac{4 - 3\beta}{\beta}, \tag{13}$$

and

$$\Gamma_1 = \frac{32 - 24\beta - 3\beta^2}{24 - 21\beta}, \tag{14}$$

where

$$\beta = \frac{p_{gas}}{p_{gas} + p_{rad}}. \tag{15}$$

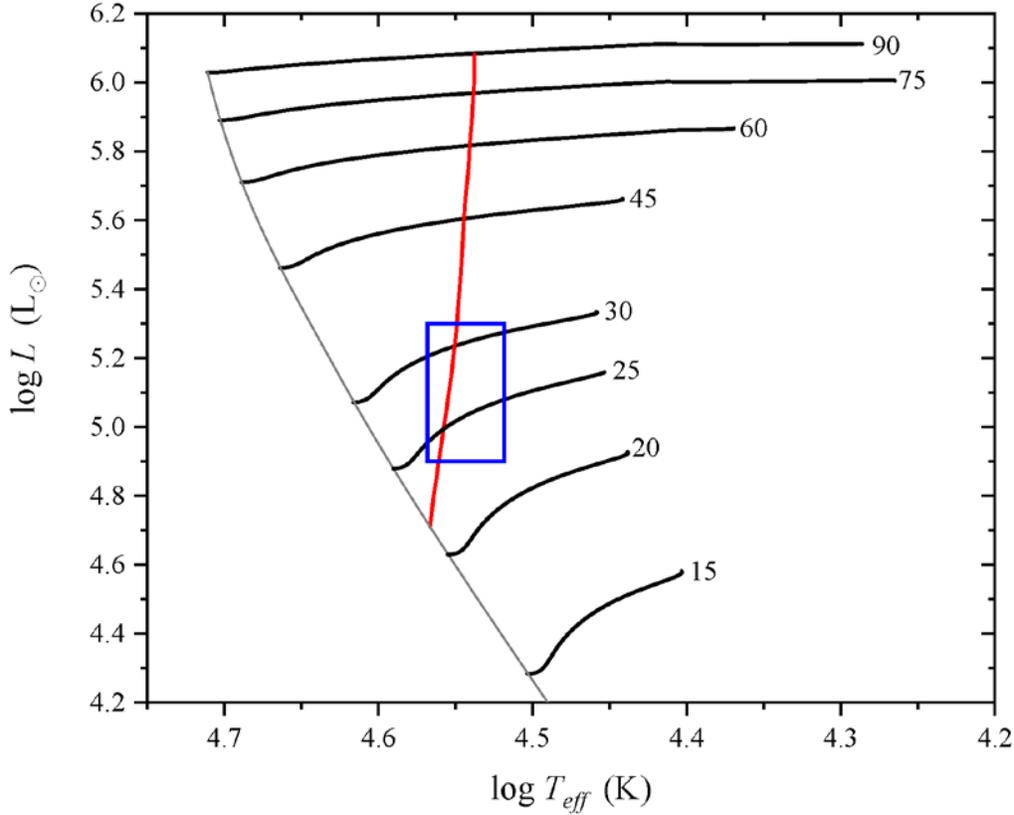

Figure 1. Evolutionary tracks in the HRD for the main sequence phase of stellar models of masses 15 to 90 $M_\odot$. The gray line shows the location of the ZAMS. Models to the right of the red line have surface $He^+$ convection zones. The blue box is the location of the strongly magnetized star NGC 1624 – 2.

## 3 CRITICAL FIELD ESTIMATES FROM STELLAR MODELS

To be able to apply the criteria in equations (3) and (11) to determine the magnetic field strength needed to suppress convection in O stars, we evolve solar composition models of masses 15 to 90 $M_\odot$ to the Terminal Age Main Sequence (TAMS), which we take to occur when the model reaches its maximum radius before hydrogen is exhausted at its center. We include mass loss during the main sequence evolution by using the mass loss rates from Vink et al. (1999, 2000, 2001). We have not included any reduction in mass loss rates that arises from the magnetic field due to closed field lines (Petit et al. 2017).

In figure 1, we show the evolutionary tracks in the Hertzsprung – Russell diagram (HRD) from the Zero Age Main Sequence (ZAMS) to the TAMS for models which do not include any evolutionary effects from potential suppression of convection.

### 3.1 Critical fields needed to suppress convection due to $He^+$ ionization

Models with $T_{eff} \lesssim 35,000$ K have convection zones due to $He^+$ ionization in addition to the Fe – opacity bump convection zone centered at $T \approx 160,000$ K. The red line in figure 1 separates models that have a He$^+$ convection zone from those that do not. O star models of mass less than ~21 M$_\odot$ have a He$^+$ convection zone at all points of their main sequence evolution. We also show in figure 1, the location of NGC 1624 – 2 (Wade et al. 2012). Clearly, this location necessitates consideration of whether the He$^+$ convection zone contributes to macroturbulence. The most developed He+ convection zone will occur in the lowest $T_{eff}$ models that are consistent with the observational constraints ($T_{eff}$ = 33000 K for $M$ = 22 – 31 M$_\odot$). Because the temperature at the center of the convection zone (~38,000 K) is only slightly higher than the photospheric temperature, the sound speed and pressure are very low compared to their values in the Fe – bump convection zone. Much smaller fields are thus needed to suppress the He+ convection. The Sundqvist criterion indicates, depending on the stellar mass, that a total field greater than 280 - 380 G is sufficient to suppress the He+ convection, whereas the modified Gough-Tayler criterion requires a vertical field greater than 120 - 160 G. These fields are smaller than found in most magnetic O stars (see S13) and so the He$^+$ convection zone is predicted to be suppressed in such stars and excludes the possibility that they generate macroturbulence.

**3.2 Critical fields needed to suppress convection due to the Fe opacity bump**

In figure 2, we compare estimates of the critical vertical field needed to suppress convection associated with the Fe opacity bump in ZAMS stars. We apply the criteria at all points in the Fe-Bump convection zone.

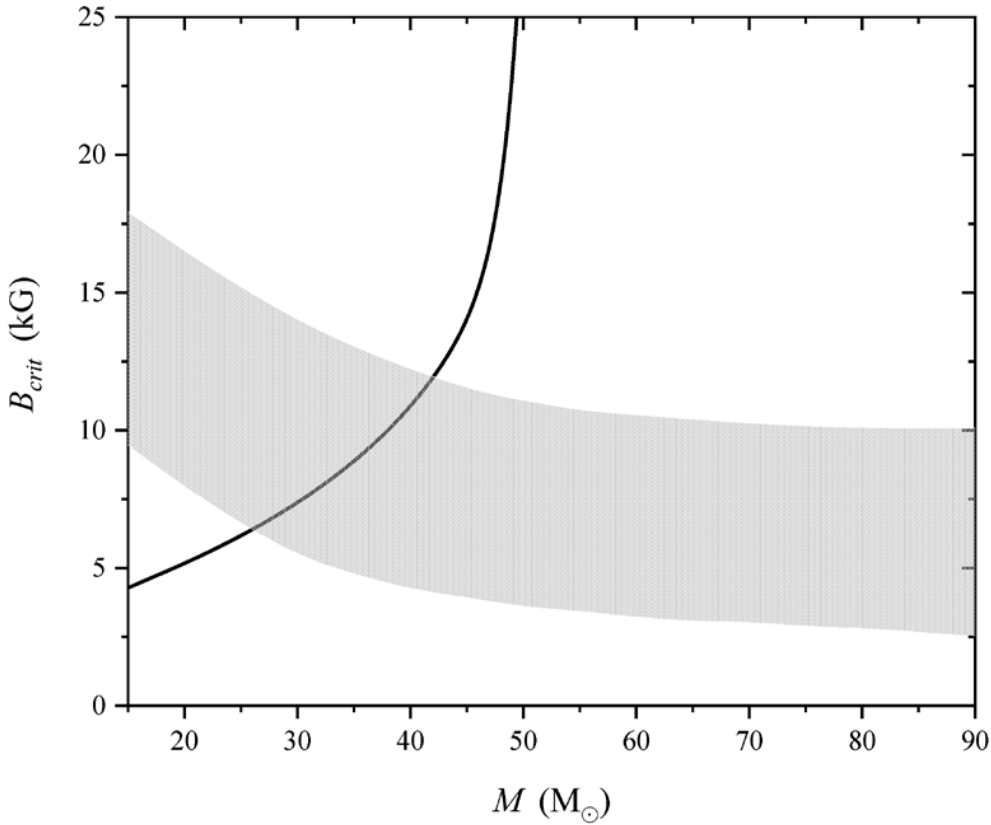

Figure 2. Critical field, $B_{crit}$, needed to suppress Fe-bump convection in ZAMS models is plotted as a function of stellar mass, $M$. The black line is for predictions of the vertical field strength from the modified Gough-Tayler criterion. The gray band is for predictions of the total field strength based on equating the thermal and magnetic pressures at each point of the convection zone.

In the ZAMS models, the surface convection zones carry only a small fraction of the energy flux (less than 2%) and magnetic suppression of the convection zone has a very small effect on the stellar structure, which makes the non-magnetic models suitable for estimating the critical field strengths. The two criteria make similar predictions for masses near 30 $M_\odot$, but differ significantly at other masses. This difference is a consequence of the increasing relative importance of radiation pressure to gas pressure with mass. For $M \lesssim 30$ $M_\odot$, gas pressure dominates and $Q \approx 1$. For $M \gtrsim 30\,M_\odot$, radiation pressure dominates and $Q \gg 1$. As noted above, because of the increased buoyancy in radiation pressure dominated envelopes of massive stars, there **is** a limit on the mass for which magnetic field can suppress convection. For ZAMS stars, we find this limit is between 53 and 54 $M_\odot$.

Unlike the ZAMS models, the fraction of the energy flux that is carried by convection can be significant for the TAMS models. For TAMS models of mass 15 $M_\odot$, 5% of the energy is carried by convection but this increases to 44% at 90 $M_\odot$. For the higher masses, magnetic suppression of convection is expected to have a significant effect on stellar structure. We begin by using our non-magnetic models to estimate the critical fields needed to suppress Fe-bump convection in TAMS stars. These critical field estimates are shown in Figure 3.

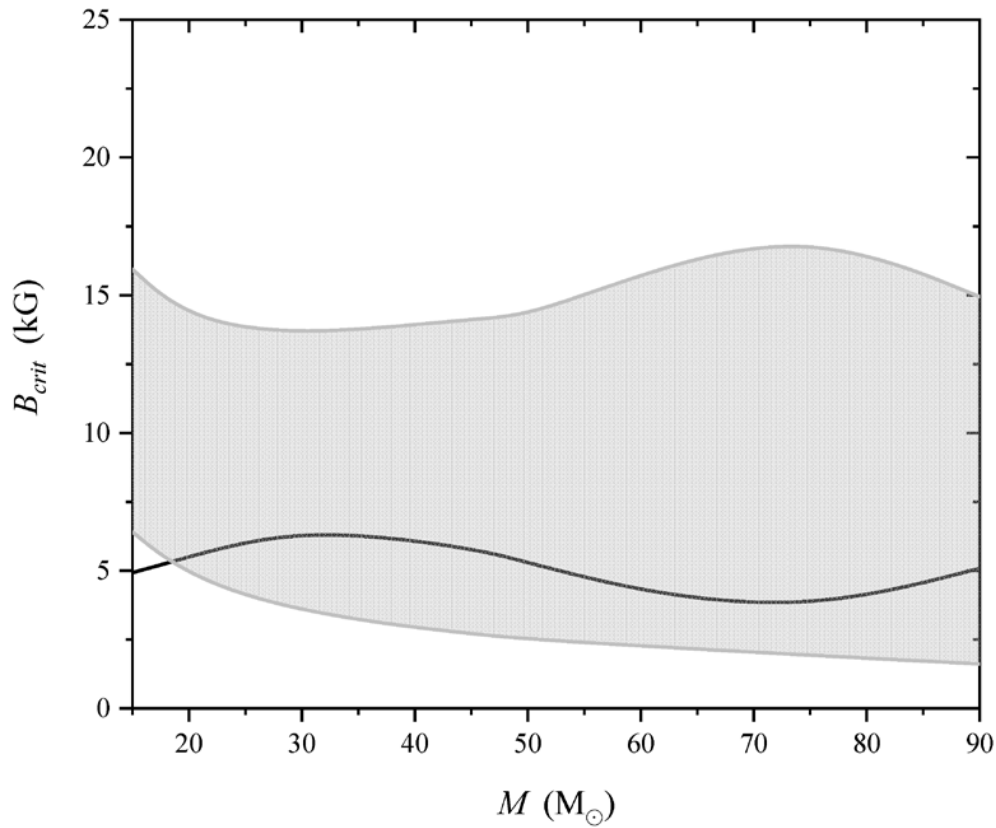

Figure 3. Critical field, $B_{crit}$, needed to suppress Fe-bump convection in TAMS models is plotted as a function of stellar mass, $M$. The black line is for predictions of the vertical field strength from the modified Gough-Tayler criterion. The gray band is for predictions of the total field strength based on equating the thermal and magnetic pressures at each point of the convection zone.

## 3.4 Stellar evolution models with magnetic inhibition of convection

To investigate how magnetic suppression of convection affects the stellar structure, we have calculated models with the usual Schwarzschild criterion for convection, $\nabla > \nabla_{ad}$, replaced by one consistent with equation (11), i.e.

$$\nabla > \nabla_{ad} + \delta, \tag{16}$$

where

$$\delta = \frac{1}{Q} \frac{v_A^2}{v_A^2 + c_s^2}. \tag{17}$$

In these models, the magnetic field is assumed vertical and its strength is assumed not to vary with depth. This is appropriate because of the small thickness of the Fe-opacity bump convective zone. To include the suppressing effects of the magnetic field in the mixing length prescription for convective energy transport, we simply replace $\nabla - \nabla_{ad}$ by $\nabla - \nabla_{ad} - \delta$.

We have calculated models for masses from 15 to 90 M$_\odot$ in increments of 15 M$_\odot$ and vertical fields 0 to 20 kG in increments of 4 kG. We impose the magnetic field on the pre-main sequence and evolve the model to the ZAMS without mass loss. Mass loss is 'turned on' at the ZAMS and the model is then evolved to the TAMS. The field strength is kept constant throughout the evolution, which simplifies attaining the goal of our calculations, i.e. to determine in a self-consistent way the critical field needed to suppress convection in a stellar model at a particular point of its main sequence evolution.
Bearing in mind that the modified Gough-Tayler criterion only involves the vertical component of the magnetic field, we see that the two criteria give similar values for the critical field needed to suppress convection in TAMS models. We also see that the modified Gough-Tayler criterion allows the possibility of magnetic suppression for all of the mass range considered here. The reason for this is that the TAMS models are more luminous than their ZAMS counterparts and, because convection has to carry a greater fraction of the energy flux, convection is more efficient in the TAMS models. As a result, $\nabla - \nabla_{ad}$ is smaller in the TAMS models and $Q(\nabla - \nabla_{ad})$ remains below unity. However, because the magnetic field inhibits convective energy transport, the convective efficiency decreases with increasing vertical field strength and $Q(\nabla - \nabla_{ad})$ increases, possibly in excess of unity. Thus, the feedback of magnetic inhibition effects on the stellar structure must be taken into account for models in which convection carries a significant fraction of the energy flux.

The evolutionary tracks in the HRD are shown in figure 4. The black and orange circles show where the stellar age is equal to 25%, 50% and 75% of the main sequence lifetime for field strengths 0 and 20 kG, respectively. We see that for the 30 and 45 M$_\odot$ models (top row), the magnetic field does not significantly alter the evolution compared to non-magnetic stars of the same mass, even though the higher magnetic field (20kG, orange curve) is able to suppress convection for most of the main sequence lifetime. For the higher mass models, the evolution can be significantly altered by the magnetic field in the late stages of main sequence evolution. We stress here that in these models the field strength has been kept fixed. Depending on the interior magnetic field profile, it is possible that the field decreases in strength as the star expands, assuming that magnetic flux is conserved. If this is the case, the evolutionary impact of magnetic fields could be less than shown in figure 4.

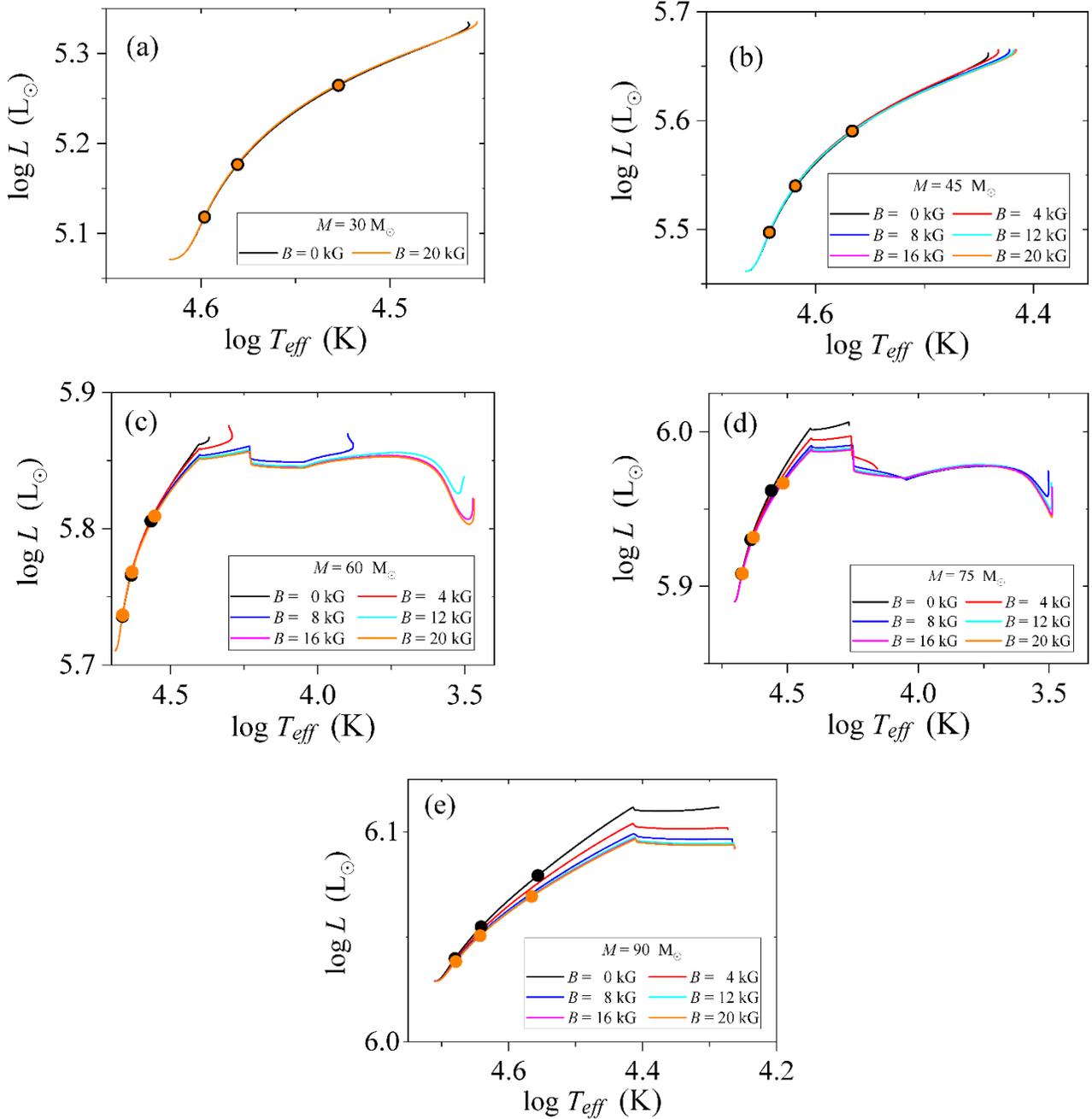

Figure 4. Evolutionary tracks in the HRD from ZAMS to TAMS. Each panel contains a number of tracks for a single mass, $M$. The tracks differ in the adopted magnetic field strength, $B$. The black and orange circles show where the stellar age is equal to 25%, 50% and 75% of the main sequence lifetime for field strengths 0 and 20 kG, respectively. For the $M = 30$ and 45 $M_\odot$ tracks the presence of the magnetic field has little effect because convective energy transport is inefficient on all of the main sequence. For the more massive models, the efficiency of convection increases in the late stages of main sequence evolution and so the evolution depends significantly on the field strength.

The evolutionary models verify our estimates of the field strengths needed to suppress Fe-opacity bump convection in ZAMS models of mass less than ~ 55 $M_\odot$ and that Fe-opacity bump convection cannot be magnetically suppressed for masses greater than ~ 55 $M_\odot$. For the TAMS models, we find that the reduction of convective efficiency due to the magnetic field causes the star to expand as it does for lower main sequence stars (Mullan & MacDonald 2001). To illustrate this finding, we show in figure 5 how the radii of 75 $M_\odot$ models change with time for a range of magnetic field values. The curves for $B$ = 12, 16, 20 kG almost coincide because although the magnetic field does not completely suppress convection it does make convective energy transport inefficient.

The evolution of the stars to the red, associated with their expansion, causes the location of the Fe-opacity bump to move deeper into the star. The increase in thermal pressure reduces $\delta$ and the magnetic inhibition of convection becomes reduced.

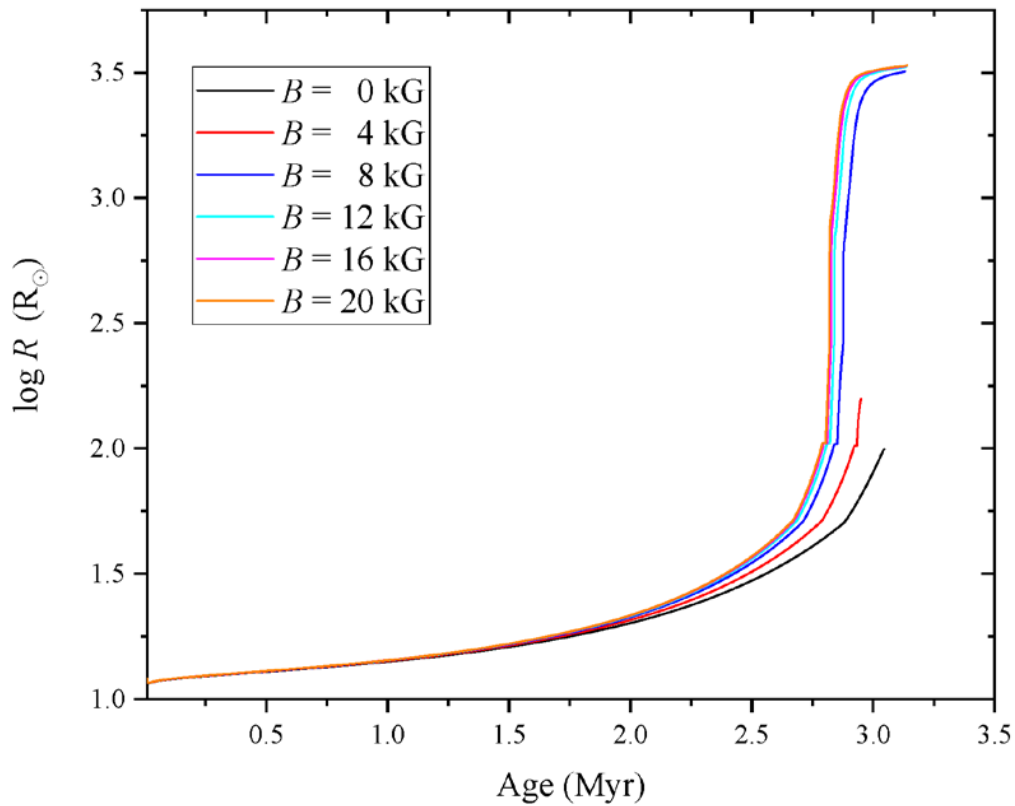

Figure 5. Evolution of the stellar radius for models of mass 75 $M_\odot$ and a range of magnetic field strengths as show in the legend.

The minimum vertical field strength to suppress convection for a stellar model at a given point in the HRD is shown in figure 6. This plot was constructed by determining the point on an evolutionary track for given mass and field strength at which convection due to the Fe-opacity bump is just suppressed, i.e. at this point there is a transition between being convective and fully radiative. We find that an upper mass limit for complete magnetic suppression occurs at all points on the main sequence. This limit is shown by the magenta line in figure 6. The upper mass limit decreases from ~ 55 $M_\odot$ on the ZAMS to ~30 $M_\odot$ on the TAMS. The reduction of the upper mass limit as the stars evolve from ZAMS to TAMS is a consequence of the location of the Fe-opacity bump moving deeper into the star.

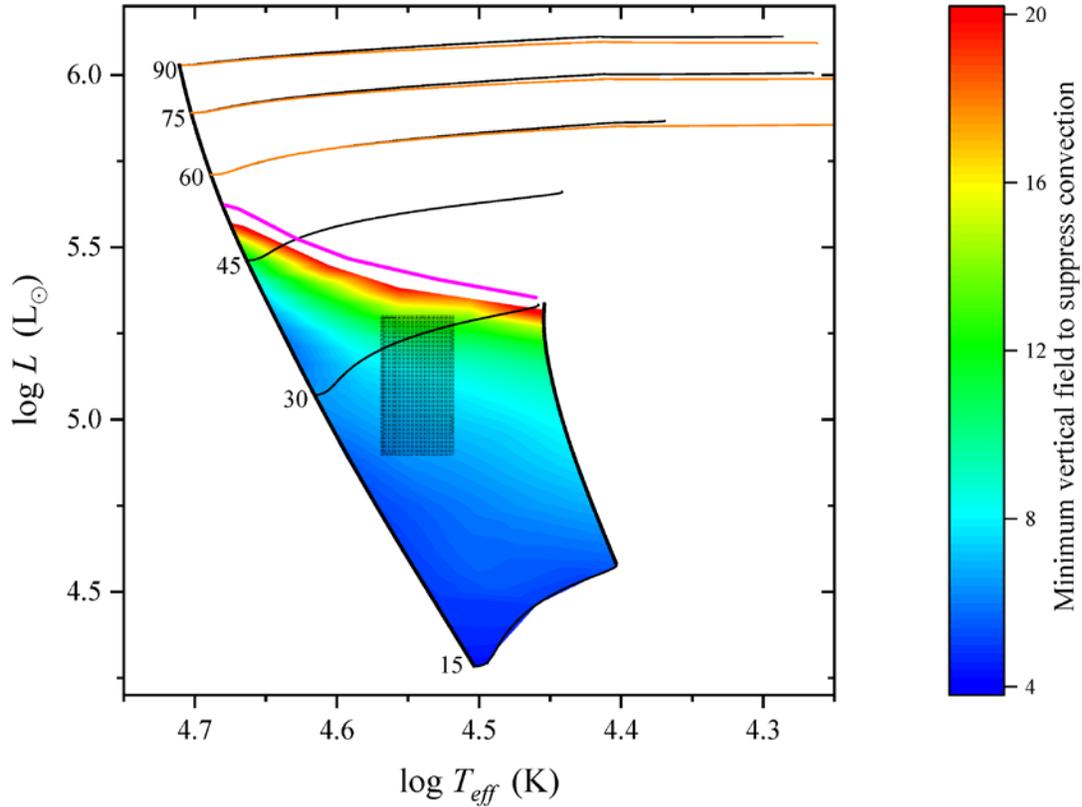

Figure 6. Color plot of minimum vertical field (in kG) needed to suppress Fe – opacity bump convection. The magenta line shows the upper limit for magnetic suppression to be possible. The thick black lines show the locations of the ZAMS and TAMS. Also shown are evolutionary tracks for non-magnetic models (black lines) and models with $B = 20$ kG (orange lines). The tracks for the 60 and 75 $M_\odot$ magnetic models have been truncated at the right axis. The rectangle shows the location of NGC 1624 – 2 according to Wade et al. (2012).

## 4 CONCLUSIONS AND DISCUSSION

We have derived a condition for convective instability in the presence of a vertical magnetic field for a general equation of state that includes radiation pressure and other non-ideal processes. Our criterion is a generalization of the Gough & Tayler (1966) condition for an ideal gas. We have used our criterion to investigate magnetic suppression of convection associated with the Fe-opacity bump in massive stars.

    We find that an upper mass limit for magnetic suppression occurs at all points on the main sequence. This upper mass limit decreases from ~55 $M_\odot$ on the ZAMS to ~30 $M_\odot$ on the TAMS.

    We show that use of non-magnetic models for estimating critical suppression field strengths is a valid approximation for ZAMS stars in which only a small fraction of the energy flux is carried by convection but is not a good approximation for more evolved stars in which a significant fraction of the energy flux is carried by convection. For evolved stars, magnetic suppression of convection can significantly alter the stars' evolution.

    For NGC 1624 – 2, we find the minimum vertical field required to suppress convection is 9.5 ± 3.4 kG. If we assume that value corresponds to the root mean square average of the vertical component of a dipole, the polar dipole strength is 16.5 ± 5.9 kG. For this star, Wade et al. (2012) detected a 5.35 ± 0.5

kG longitudinal magnetic field and probable Zeeman splitting of the Stokes *I* profiles of metal lines corresponding to a surface field $\langle B \rangle = 14 \pm 1$ kG, and a dipole of surface polar strength ∼20 kG. The 30 $M_\odot$ track indicates that the complete suppression of convection should never occur over the MS lifetime if the field strength were to remain constant.

The suppression of macro-turbulence, if and when driven solely by convection, in magnetic massive stars may provide a way to identify very strong fields spectroscopically. As magnetic O-type stars tend to rotate very slowly, macroturbulence is the main line broadening-mechanism (S13). With its suppression, very magnetic massive stars should display very narrow spectral lines that would also help to detect Zeeman splitting. However, our results suggest that this would not be the case for very massive stars for which convection is not completely suppressed even for very large fields. For typical macroturbulence velocities of 50 km/s, Zeeman splitting would only become dominant in terms of spectral line shape for field strengths in excess of 50-100 kG.

## ACKNOWLEDGEMENTS

This work is supported in part by the Delaware Space Grant (JM). VP acknowledges support by the National Science Foundation under Grant No. 1747658 and the University of Delaware Research Foundation.

# APPENDIX. MAGNETIC INHIBITION OF CONVECTION IN RADIATION DOMINATED ENVELOPES

Assuming ideal MHD, the evolution of the magnetic field is given by the induction equation

$$\frac{\partial \mathbf{B}}{\partial t} = \nabla \times (\mathbf{v} \times \mathbf{B}). \tag{18}$$

The equation of motion for the velocity field is

$$\rho \frac{d\mathbf{v}}{dt} = -\nabla p + \rho \mathbf{g} + \frac{1}{4\pi}(\nabla \times \mathbf{B}) \times \mathbf{B}. \tag{19}$$

The continuity equation is

$$\frac{d\rho}{dt} + \rho \nabla \cdot \mathbf{v} = 0. \tag{20}$$

In the absence of heat transfer, thermodynamic changes are adiabatic so that

$$\frac{d \ln p}{dt} = \Gamma_1 \frac{d \ln \rho}{dt}. \tag{21}$$

For an initially uniform vertical field, the evolution of small perturbations, denoted by subscript 1, to an initially static state, denoted by subscript 0, is determined by the following equations:

The magnetic field equation is

$$\frac{\partial \mathbf{B}_1}{\partial t} = \nabla \times (\mathbf{v} \times \mathbf{B}_0) = (\mathbf{B}_0 \cdot \nabla) \mathbf{v} - \mathbf{B}_0 \nabla \cdot \mathbf{v}. \tag{22}$$

The equation of motion is

$$\rho_0 \frac{\partial \mathbf{v}}{\partial t} = -\nabla p_1 + \rho_1 \mathbf{g} + \frac{1}{4\pi}(\nabla \times \mathbf{B}_1) \times \mathbf{B}_0. \tag{23}$$

The continuity equation is

$$\frac{\partial \rho_1}{\partial t} + \nabla \cdot (\rho_0 \mathbf{v}) = 0. \tag{24}$$

The thermal energy equation is

$$\frac{\partial p_1}{\partial t} + \mathbf{v} \cdot \nabla p_0 = c_s^2 \left( \frac{\partial \rho_1}{\partial t} + \mathbf{v} \cdot \nabla \rho_0 \right), \tag{25}$$

where $c_s$ is the adiabatic sound speed.

Following MIS89, we take the time derivative of equation (23) and use equations (24), (25) and (22) to eliminate the time derivatives of $\rho_1$, $p_1$, and $\mathbf{B}_1$. Then

$$\rho_0 \frac{\partial^2 \mathbf{v}}{\partial t^2} = \nabla \left( c_s^2 \rho_0 \nabla \cdot \mathbf{v} + \mathbf{v} \cdot \nabla p_0 \right) - \nabla \cdot (\rho_0 \mathbf{v}) \mathbf{g} + \frac{1}{4\pi} \left( \nabla \times \left[ \nabla \times (\mathbf{v} \times \mathbf{B}_0) \right] \right) \times \mathbf{B}_0. \tag{26}$$

Taking the z- axis to be in the up direction, and assuming perturbations of form $A(z) e^{i(\omega t + ky)}$, we find

$$\left( \omega^2 - k^2 c_s^2 - k^2 v_A^2 \right) v + v_A^2 \frac{\partial^2 v}{\partial z^2} = ik \left( gw - c_s^2 \frac{\partial w}{\partial z} \right), \tag{27}$$

and

$$\omega^2 w + \frac{1}{\rho_0} \frac{\partial}{\partial z} (c_s^2 \rho_0) \frac{\partial w}{\partial z} + c_s^2 \frac{\partial^2 w}{\partial z^2} = -ik \left[ \frac{1}{\rho_0} \frac{\partial}{\partial z} (c_s^2 \rho_0) v + gv + c_s^2 \frac{\partial v}{\partial z} \right], \tag{28}$$

where

$$\mathbf{v} = (u, v, w). \tag{29}$$

MIS89 show that to investigate stability only modes with horizontal wavelength much smaller that the vertical scale length of any of the equilibrium quantities need to be considered (also see Lou 1989). In this limit, the perturbation equations are shown to have two solutions; a high frequency fast-mode solution and a slow wave solution of finite frequency that determines the stability of the equilibrium state.

Following Syrovatskii & Zugzhda (1967), MIS89 make the approximations

$$\omega^2 \ll \left( c_s^2 + v_A^2 \right) k^2 \tag{30}$$

and

$$k^2 H_z^2 \gg 1 \tag{31}$$

to isolate the slow wave. Equation (27) then becomes

$$\left( c_s^2 + v_A^2 \right) k^2 v = ikc_s^2 \frac{\partial w}{\partial z} - ikgw, \tag{32}$$

which on using to eliminate v from equation (28) gives

$$c_T^2 \frac{\partial^2 w}{\partial z^2} + \left[ \frac{v_A^2}{v_f^2} \frac{1}{\rho_0} \frac{\partial}{\partial z} (c_s^2 \rho_0) - c_s^2 \frac{\partial}{\partial z} \left( \frac{c_s^2}{v_f^2} \right) \right] \frac{\partial w}{\partial z} + \left\{ \omega^2 + \frac{g}{v_f^2} \left[ \frac{1}{\rho_0} \frac{\partial}{\partial z} (c_s^2 \rho_0) + g \right] + c_s^2 g \frac{\partial}{\partial z} \left( \frac{1}{v_f^2} \right) \right\} w = 0,$$
$$\tag{33}$$

where

$$c_T^2 = \frac{c_s^2 v_A^2}{c_s^2 + v_A^2}, \quad v_f^2 = c_s^2 + v_A^2. \tag{34}$$

MIS89 then show that there are no unstable solutions if the coefficient of $w$ in equation (33) is negative. Since instability requires that $\omega^2 < 0$, it follows that there are no unstable solutions if

$$\frac{1}{v_f^2}\left[\frac{1}{\rho_0}\frac{d}{dz}\left(c_s^2 \rho_0\right) + g\right] + c_s^2 \frac{d}{dz}\left(\frac{1}{v_f^2}\right) < 0. \tag{35}$$

For the case of an initially uniform magnetic field, evaluation of the derivative of $v_f^{-2}$ leads to

$$\frac{c_s^2}{\rho_0}\frac{d\rho_0}{dz} + g + \frac{v_A^2}{v_f^2}\frac{1}{\rho_0}\frac{d\left(\rho_0 c_s^2\right)}{dz} < 0. \tag{36}$$

The criterion in equation (36) is quite general but not in terms typically used for convection criteria. By multiplying throughout by

$$H_p = -\frac{dz}{d\ln p_0} = \frac{p_0}{\rho_0 g}, \tag{37}$$

equation (36) can be written as

$$-\frac{d\ln \rho_0}{d\ln p_0} + \frac{1}{\Gamma_1} - \frac{v_A^2}{v_f^2}\left(1 + \frac{d\ln \Gamma_1}{d\ln p_0}\right) < 0. \tag{38}$$

By relating $d\ln \rho_0 / d\ln p_0$ to the structural gradient and $\Gamma_1^{-1}$ to the adiabatic gradient, equation (38) leads to the stability criterion

$$Q(\nabla - \nabla_{ad}) - \frac{v_A^2}{v_f^2}\left(1 + \frac{d\ln \Gamma}{d\ln p_0}\right) < 0. \tag{39}$$

This is the same as the criterion in equation (11), proposed by MacDonald & Mullan (2009), provided $\Gamma_1$ does not vary with depth.

15